\documentclass{svjour3}                     
\smartqed  
\usepackage{graphicx}
\usepackage{amsmath,amssymb}
\usepackage{bbm}

\newcommand{\openone}{\mathbbm{1}}

\newcommand{\eps}{\varepsilon}
\newcommand{\vphi}{\varphi}
\newcommand{\Op}[1]{\boldsymbol{\mathsf{\hat{#1}}}}

\newcommand{\Fkt}[1]{\,\mathsf {#1}}

\ifx\Tr\renewcommand{\Tr}{\Fkt{Tr}}
\else\newcommand{\Tr}{\Fkt{Tr}}
\fi

\renewcommand{\dagger}{+}

\newcommand{\uzero}{c_3}
\newcommand{\rydbergBra}{\langle r |}
\newcommand{\rydbergKet}{| r \rangle}
\newcommand{\rrKet}{| r r \rangle}

\journalname{Quantum Information with Neutral Particles}
\begin{document}

\title{Prospects for fast Rydberg gates on an atom chip}
\author{Matthias M. M\"uller         \and
  Harald R. Haakh \and Tommaso Calarco \and Christiane P. Koch \and
  Carsten Henkel
}

\authorrunning{M\"uller, Haakh, Calarco, Koch, Henkel} 

\institute{Matthias M. M\"uller, Tommaso Calarco \at
  Institut f\"ur Quanteninformationsverarbeitung, 
  Universit\"at Ulm, 89081 Ulm, Germany
  \and
  Harald R. Haakh, Carsten Henkel \at
  Institut f\"ur Physik und Astronomie, Universit\"at Potsdam, 
  Karl-Liebknecht-Str. 24/25, 14476 Potsdam, Germany
  \and
  Christiane P. Koch \at
  Institut f\"ur Physik, Universit\"at Kassel, 
  Heinrich-Plett-Str. 40, 34132 Kassel, Germany\\
  \email{christiane.koch@uni-kassel.de}           
}

\date{Received: date / Accepted: date}

\maketitle

\begin{abstract}
Atom chips are a promising candidate for a scalable architecture for
quantum information processing provided a universal set of gates can
be implemented with high 
fidelity. The difficult part in achieving universality is the
entangling two-qubit gate. We consider a Rydberg phase gate for two
atoms trapped on a chip and employ optimal control theory to find the
shortest gate that still yields a reasonable gate error. Our parameters
correspond to a situation where the Rydberg blockade regime is not yet
reached. We discuss the role of spontaneous emission and
the effect of noise from the chip surface on
the atoms in the Rydberg state.
%
\end{abstract}

\section{Introduction}
\label{sec:intro}
Neutral trapped atoms allow for qubit encoding in metastable internal
states; they offer long decoherence times and can easily be
controlled by optical and magnetic fields. This makes them a
promising candidate for the realization of
a quantum computer~\cite{ZollerEPJD05}. 
Since the atoms hardly interact with each other, 
the largest difficulty encountered when building such a
quantum computer is the implementation of an entangling two-qubit
operation.
The clock speed is thus typically limited by 
the gate operation time for one two-qubit gate which, together with
several single-qubit operations, provides universal quantum
computing~\cite{NielsenChuang}.  

A number of schemes to entangle qubits carried by neutral atoms have
been proposed~\cite{BrennenPRL99,JakschPRL99,CalarcoPRA00,JakschPRL00,BuchkremerLP02,SoerensenPRL04,NegrettiEPJD05,TreutleinPRA06,MuellerPRL09}.
An estimate of the corresponding gate duration is obtained in terms of
the inverse of the interaction strength promoting the entanglement.
Adiabatic time evolution is not essential; it can be avoided by
employing optimal control
theory~\cite{SomloiCP93,ZhuJCP98,PalaoPRL02,TeschPRL02,PalaoPRA03}. 
In fact, optimal control theory is an extremely versatile tool that
allows for achieving a high-fidelity implementation of desired
quantum tasks. Moreover, it can be used to determine what
fundamentally limits fidelity and duration of a quantum
operation~\cite{TommasoPRL09}. 

For a two-qubit gate with neutral atoms, 
the highest gate speeds can be expected when the atoms are internally
excited to expose them to long-range interactions, 
a prominent example being dipole-dipole forces between Rydberg 
atoms~\cite{JakschPRL00,IsenhowerPRL10,WilkPRL10,SaffmanRMP10}. However, care needs
to be taken if the atoms are \textit{resonantly} excited into a state
with long-range interactions. 
For the example of the Rydberg gate, this will happen, e.g., if a perfectly
entangling gate is desired but the distance
between the atoms is too large to reach the Rydberg blockade regime
\cite{MMMueller}. 
The forces that induce entanglement will
then couple internal electronic and vibrational dynamics, and 
the motional state of the atoms will have changed after the gate
operation, implying strong leakage out of the quantum register.
In principle, the vibrational excitation energy can be absorbed by 
an external field determined for example by optimal control
theory. However, this requires the gate duration to be long enough to
resolve the motional energy levels~\cite{Goerz}. For resonant
excitation into states with long-range interaction, the minimum
duration, or quantum speed limit,
 for an entangling gate is thus limited either by the inverse
interaction strength or by the vibrational period of the trap,
whichever one is longer~\cite{MMMueller,Goerz}.

Trapping, controlling and entangling two atoms is just the very first
step towards a neutral atom based quantum computer. 
A scalable architecture is required to assemble many qubits and carry
out any meaningful computation. One possibility to
achieve scalability is given by 
miniaturizing the tools for trapping and
manipulating the atoms to fit on a micrometer-sized 
chip~\cite{FolmanAMOP02,FortaghRMP07,ReichelVuleticBook}. These
microtraps offer strong confinement (fast vibrational motion) that may
help in fast entangling gates. 
However, the close proximity of the atoms to the surface of the chip
introduces noise sources which might significantly compromise the gate
performance.
Here, we study whether the fastest possible entangling
gates for neutral atoms known to date -- optimal Rydberg phase gates at
the quantum speed limit -- can be implemented on an atom chip, despite
the presence of noise. 
To this end, we present an optimal control analysis for a pulsed excitation
scheme of two atoms, using trap parameters realistic for microtraps. 
The analysis includes the nuclear
motion of the atom pair and addresses the influence of spontaneous
emission from electronically excited states. We also estimate the influence
of electric fields (static and dynamic) that arise from surface contamination
and from blackbody and thermal near field radiation. 
Attention is payed to the specific
properties of the fragile, highly excited Rydberg states that show huge
electric dipole moments. Our results suggest that a fast gate (on the
scale of some 10\,ns) is possible with errors on the level of $10^{-3}$.
We identify options for reducing this error by another order
of magnitude to reach the fault tolerance threshold.

\section{Fast phase gates via optimal control}
\label{sec:oct}

\subsection{Controlled phase gate via Rydberg interaction of neutral atoms}
\label{subsec:rydberg}
\begin{figure}[bt]
  \includegraphics*[width=0.5\linewidth]{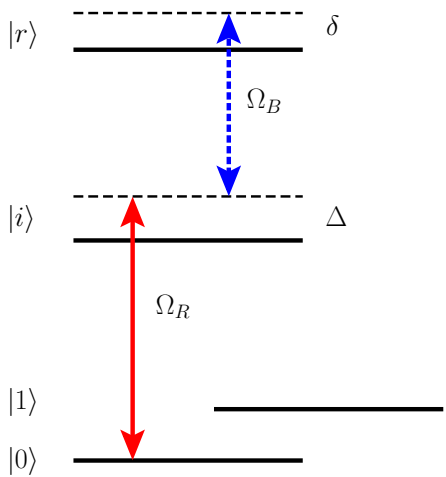}
\caption{One-atom level scheme for near-resonant two-photon
  excitation to a Rydberg state.
  \label{Fig:OneAtomHam}
  }
\end{figure}

We consider two rubidium atoms that are trapped with a typical
distance of $4\,\mu$m between them. The atoms sit in the ground
state of the trap which is assumed to be deep enough such that each
well can be approximated by a harmonic potential. The qubit states are
encoded in two hyperfine levels of the ground state, for example 
$|0\rangle=|5s_{1/2},F=2,M_F=2\rangle$, 
$|1\rangle=|5s_{1/2},F=1,M_F=1\rangle$.
A controlled phase gate can be implemented by excitation to a Rydberg
state where the two atoms are exposed to a long-range
interaction~\cite{JakschPRL00}. In a recent experiment with two
rubidium atoms held in optical tweezers~\cite{GaetanNatPhys09},  
the $\rydbergKet=|58{\rm d}_{3/2},F=3,M_F=3\rangle$
Rydberg state was accessed by a near-resonant two-photon transition
via the $|i\rangle=|5{\rm p}_{1/2},F=2,M_F=2\rangle$ intermediate state.
The Hamiltonian for a single trapped atom in the rotating-wave
approximation reads 
\begin{eqnarray}
  \label{eq:H1atom}
  \Op{H}^{(1)}(t) &=& 
  \sum_{j=0,1} |j\rangle\langle j| \otimes
  \left(\Op{T}_{\Op{r}} + V^{j}_{trap}(\Op{r})\right) \nonumber \\
  &&+ |i\rangle\langle i| \otimes 
  \left(\Op{T}_{\Op{r}} + V^{i}_{trap}(\Op{r}) +
    \frac{\Delta}{2}\right) 
  + \rydbergKet\rydbergBra \otimes
  \left(\Op{T}_{\Op{r}} + V^{r}_{trap}(\Op{r}) +
    \frac{\delta}{2}\right) \nonumber \\
  &&+ \frac{\Omega_R(t)}{2} 
  \bigg( |0\rangle\langle i| + |i\rangle\langle  0 |\bigg) \otimes
  \openone(\Op{r})   + \frac{\Omega_B(t)}{2} 
  \bigg( |i\rangle\rydbergBra + \rydbergKet\langle i |\bigg) \otimes \openone(\Op{r})\,.
\end{eqnarray}
Here, $\Op{r}$ denotes the position operator of the atom, 
$\Op{T}$
the kinetic energy operator, and $\Omega_\lambda(t)$ ($\lambda = R, \, B$)
the
time-dependent Rabi frequencies of the red and blue lasers
(wavelengths $795\,$nm and $474\,$nm for the chosen Rydberg state).
The Rabi frequencies are parametrized according to 
$\Omega_\lambda(t)=\Omega_{j,0}(\tanh\eps_\lambda(t)+1)/2
\,\in\,[0,\Omega_{j,0}]$,
where $\eps_\lambda(t)$ will be determined by optimal control.
$\Delta$ denotes the detuning of the red laser with respect to the
intermediate state. 
The two-photon detuning from the
Rydberg level is given in terms of the Stark shift,
$\delta=(\Omega_{B,0}^2-\Omega_{R,0}^2)/4\Delta$. 
The corresponding level scheme is depicted in
Fig.~\ref{Fig:OneAtomHam}. 
The electronically excited states $|i\rangle$ and $\rydbergKet$
decay via spontaneous emission, their
lifetimes are $27.7\,$ns for $|i\rangle$ and $210\,\mu$s for
$\rydbergKet$ (see also Sec.~\ref{s:cavity-QED}).
The total Hamiltonian for the two atoms is given by
\begin{equation}
  \label{eq:H2atom}
  \Op{H} = \Op{H}^{(1)}_1 \otimes \openone_{4,2} \otimes
  \openone_{\Op{r}_2} + \openone_{4,1} \otimes
  \openone_{\Op{r}_1} \otimes \Op{H}^{(1)}_2 + \Op{H}^{(1,2)}_{int}
\end{equation}
with the interaction term describing the long-range forces when both
atoms are in the Rydberg state, 
\begin{equation}
  \label{eq:Hint}
  \Op{H}^{(1,2)}_{int}= \vert rr\rangle\langle rr\vert 
     \otimes \frac{\uzero}{|\Op{r}_{1}-\Op{r}_{2}|^3}\,.
\end{equation}
The parameter $\uzero = 3.433 \cdot 10^6\,{\rm a.u.} =
3230 \,{\rm MHz}\,\mu{\rm m}^3$ has been calculated in
Ref.~\cite{GaetanNatPhys09} for the Rydberg levels $\rydbergKet$ 
and scales with the principal quantum number as $n^4$~\cite{SaffmanRMP10}.
For an interatomic distance of $r_0 = 4\,\mu$m which is realistic for
an experimental implementation,
this yields an estimated interaction energy 
of 50$\,$MHz. 
In all other states, the interaction between atoms that are
thus far apart is so weak that it can be neglected.
The harmonic approximation for the trapping potentials permits us to
integrate over the center-of-mass motion of the two atoms and to
focus on a one-dimensional model along the internuclear axis. 
We take an effective trap frequency
$\omega/2\pi \approx 276\,\mathrm{kHz}$ (period $3.6\,\mu\mathrm{s}$)
for the interatomic distance coordinate
$\Op{r} = |\Op{r}_{1}-\Op{r}_{2}|$ (ground state size 
$\approx 20\,{\rm nm}$).
%
This value is slightly larger than in the optical tweezer traps of
Ref.~\cite{GaetanNatPhys09}, and may also be achieved in magnetic
microtraps implemented on an atom chip~\cite{ReichelVuleticBook}.
The Hamiltonian, Eq.~(\ref{eq:H2atom}), is represented on an
equidistant Fourier grid extending for $\pm 0.3\,\mu$m around
$r_0=4\,\mu$m. The time evolution generated by this Hamiltonian is
obtained by solving the time-dependent Schr\"odinger
with the Chebychev propagator~\cite{RonnieReview94}. Note that
we do not attempt here to spatially resolve the Rydberg orbits of the 
excited electron in the trapping fields. The level structure in a magnetic
quadrupole field, for example, has been addressed in 
Refs.~\cite{LesanovskyEPJD05,BillPRA06}.

\subsection{Optimal control for two-qubit gates}
\label{subsec:oct}
Our goal is to implement the controlled phase gate on the qubit
register basis $\{ | 00 \rangle, | 01 \rangle, | 10 \rangle, | 11
\rangle \}$, 
\begin{equation}
  \Op{O} = \Fkt{diag} \left( e^{i \chi}, 1, 1, 1 \right)
  \quad\mathrm{or}\quad
  \Op{O} = \Fkt{diag} \left( 1, 1, 1,  e^{i \chi} \right)
  \,.
  \label{eq:target_operator}
\end{equation}
This can be achieved by finding suitable fields 
$\varepsilon = \{ \varepsilon_R(t), \varepsilon_B(t) \}$ 
that drive the system evolution from time zero to time $T$
such that $\Op{U}(T,0;\varepsilon)= \Op{O}$ up to a global phase.
The quality of the implementation is measured by the fidelity, $F$.
A suitable choice for evaluating it
is based on projecting the actual evolution
$\Op{U}(T,0;\varepsilon)$ onto the desired operation
$\Op{O}$~\cite{PalaoPRA03}, 
\begin{equation}
  \label{eq:Fre}
  F = \frac{1}{N} \left\lvert
    \Fkt{Tr}\left[\Op O^+\Op{U}(T,0;\varepsilon)\right] \right\rvert\,.
\end{equation}
$N$ denotes the number of basis states, $N=4$ if $\Op{O}$ is a
two-qubit gate. Note that the actual evolution can proceed in a
Hilbert space which is much larger than the qubit register space. In
our example, the dimension of the Hilbert space is $4\times 4\times
N_r$ where $N_r$ is the number of grid points required to represent
$\Op{r}$. Then $\Op{U}(T,0;\varepsilon)$ is obtained by a suitable
projection onto the register space which in our example is spanned by
$|\vphi_0^{jj^\prime}\rangle=
|j\rangle\otimes|j^\prime\rangle\otimes |\varphi_0\rangle$,
$j, j' = 0, 1$,
where $|\varphi_0\rangle$ represents the ground state of the trap.

Optimal control treats 
the fidelity $F$, $F \in [0,1]$, as a functional
of the control fields $\varepsilon$. Allowing for additional constraints
such as minimization of the integrated pulse energy, 
leads to the total functional $J$, 
\begin{equation}
  J = - F + \int_{0}^{T} g(\varepsilon)  dt
  ,
  \qquad
g(\varepsilon) = \frac{\alpha}{S(t)} \left[ \varepsilon(t) - \varepsilon_{\rm ref}(t)
  \right]^2,
  \label{eq:j_functional}
\end{equation}
which is to be minimized. Here $\alpha$ is a weight, 
$S(t)$ enforces the pulse to switch on and off smoothly, and
$\varepsilon_{\rm ref}(t)$ is the control field at the previous optimization
step. 
Variation of the functional $J$ with respect to
the evolving basis states and the control field yields a set of
coupled optimization equations that are solved iteratively. 
We employ here a linear variant of Krotov's
method in order to obtain a monotonically convergent
optimization algorithm~\cite{PalaoPRA03}. In
particular, the update equation for the field is then given by
\begin{equation}
  \label{eq:update}
  \Delta\varepsilon(t) = \varepsilon^{(k+1)}(t)-\varepsilon^{(k)}(t) =
  \frac{S(t)}{2\alpha}\sum_{jj^\prime}
  \langle\vphi_0^{jj^\prime}|\Op O^+ \Op U^+(t,T;\varepsilon^{(k)})\;\Op\mu\;
  \Op U(t,0;\varepsilon^{(k+1)})|\vphi_0^{jj^\prime}\rangle\,,
\end{equation}
which matches at each time $t$ 
the target states, $\Op O|\vphi_0^{jj^\prime}\rangle$,
propagated backward in time under the old fields, 
$\varepsilon^{(k)}$, with
the initial states, i.e., the basis states $|\vphi_0^{jj^\prime}\rangle$,
propagated forward in time under the new fields $\varepsilon^{(k+1)}$. 

In our example of the Rydberg gate, the short lifetime of the
intermediate state, $|i\rangle$, implies presence of a serious loss
channel. The time evolution is then not unitary
anymore. For dissipation due to spontaneous emission, the dynamics can
be modeled by a Markovian Liouville-von Neumann equation for the
density matrix with Lindblad dissipators. 
Optimal control theory can be adapted to such a
situation~\cite{AllonJCP97,OhtsukiJCP99} but the numerical effort
increases substantially. If dissipation is not needed to achieve the
control objective but it rather represents a nuisance, a simpler
approach is given by suppressing population in the dissipative
channel. This is achieved by adding an additional constraint in the
objective functional maximizing the expectation value of the projector
onto the allowed subspace
$\Op{P}_{allow} = (\openone - | i \rangle \langle i |)\otimes
(\openone - | i \rangle \langle i |)\otimes \openone_{\Op r}$,
i.e. the subspace of the total Hilbert
space excluding the unstable state~\cite{PalaoPRA08},
\begin{equation}
  J = - F + \int_{0}^{T} g(\varepsilon)  dt - 
  \int_{0}^{T} \left\langle\Op{P}_{allow}\right\rangle dt\,.
  \label{eq:j_stateconstraint}
\end{equation}
Provided the modified control objective is fulfilled, i.e., the
dissipative channel is never populated, the dynamics can again be
described by a Schr\"odinger equation. 
Including the state-dependent constraint does not modify the update
equation for the field, Eq.~(\ref{eq:update}), except for a source
term in the equation for the backward
propagation~\cite{PalaoPRA08}. Such an inhomogeneous Schr\"odinger
equation can be solved efficiently with a modified Chebychev
propagator~\cite{NdongJCP09}. 
We test the success of this scheme by propagating the initial states
under the optimal field, and adding the imaginary term
$- \frac{ {\rm i}}{ 2 } \hbar \gamma_i | i \rangle \langle i |$
to the Hamiltonian~(\ref{eq:H1atom}), describing exponential decay from 
the intermediate state $|i\rangle$.

\subsection{Limits for the duration of optimal gates}
\label{subsec:gatetime}

In the following, we employ the optimization algorithm with the standard 
functional~(\ref{eq:j_functional}) as well as with the 
functional~(\ref{eq:j_stateconstraint}) that
minimizes the population in the unstable state. We are looking for the 
best compromise between high fidelity and fast operation.

The minimum gate durations are identified by running the optimization
for varying $T$. If there is no loss, i.e., if we neglect spontaneous
emission, our system is completely controllable. In the limit of
sufficiently large $T$, the fidelity $F$ then approaches one and the gate
error approaches zero arbitrarily close~\cite{Goerz}. As $T$ is
decreased, the time spent in the Rydberg state in which the atoms
interact might become too short to pick up a non-local phase of
$\pi$ (time scale $\pi \hbar r_0^3 / \uzero \approx 10\,\mathrm{ns}$).
This is the first and most obvious constraint that limits the
gate time $T$. Secondly, when the atoms are in the Rydberg state (note
that our 
interatomic distance is too large for the blockade regime to be
reached, i.e., the atoms are resonantly excited into $\rrKet$), they
get accelerated in the attractive $1/r^3$ potential. In principle, the
optimization can find laser fields that absorb this vibrational
energy. However, in order to find such laser fields, the algorithm
needs to distinguish between the motional target state, the ground
state of the trap, and higher excited states. The vibrational period
of the trap ($\approx 3.6\,\mu\mathrm{s}$) 
may therefore also limit 
the gate operation time. For
shallow traps and a strong interaction, this second limit will be larger
than the first one, and thus determine the minimum gate time
$T$~\cite{Goerz}. 
If the interaction $\uzero/r^3$ acts over a time such
  that a phase $\pi$ is picked up, then a momentum of order
  $\delta p = 3 \pi \hbar / r_0$ is transferred to the relative motion. The motional
  error scales with the ratio to the momentum width in the ground state,
  $(\hbar \omega m/4)^{1/2}$, 
  as $1 - \exp[ - \delta p^2 / (2 \hbar \omega m ) ]
  \approx \delta p^2 / (2 \hbar \omega m)$,
see also Ref.~\cite{SoerensenPRL04}.
This translates into the condition that the ground state be much
smaller than the mean distance, 
\begin{equation}
	\sqrt{\frac{ \hbar }{ m \omega} } 
	\ll \frac{ \sqrt{ 2 } }{ 3\pi } r_0 \approx 0.15 \, r_0\,,
	\label{eq:estimated-force-kick}
\end{equation}
which is well met for our parameters since the trap ground state has 
a size $\approx 20\,{\rm nm}$.

In our calculations the trap is only used to determine the initial
state. During the action of the laser fields, the trapping potential
is set to zero which corresponds to the dipole trap being 
switched off in the experiment~\cite{GaetanNatPhys09}. We thus prevent
the optimization algorithm from finding solutions that, if necessary,
absorb vibrational energy. Once vibrational dynamics start to play a
role, this will show up as a decrease in fidelity. However, this loss of
fidelity could be avoided by leaving the trap on during the gate, in
other words, it is not caused by a fundamental limit.
  Note that the trap is only needed to define the
  desired motional state at the end of the gate for the electronic
  ground state which carries the qubit state.

Finally, a third factor limiting the gate time $T$
is due to spontaneous emission from the intermediate
state. While it turns out that this limits the best
possible fidelity in current implementations of the Rydberg
gate~\cite{GaetanNatPhys09}, 
it is not a fundamental limit since one could think of accessing the
Rydberg states differently, avoiding near-resonant two-photon
excitation via such a strongly decaying state. 
We therefore discuss gate times and errors for both cases,
including and neglecting spontaneous emission for the sake of comparison.

\subsection{Minimizing spontaneous emission: adiabatic pulse sequence}

\begin{table}[tbp]
  \centering
  \begin{tabular}{|c|cc|}
    \hline
    gate time $T$ 
    & \multicolumn{2}{c|}{gate error $\epsilon$}
    \\
    & no loss 
    & with loss 
    \\
    \hline
30ns &	0.294	&	0.305 \\
40ns &	0.043	&	0.221 \\
50ns &	0.003	&	0.025 \\
60ns &	0.003	&	0.021 \\
70ns &	0.004	&	0.021 \\
80ns &	0.005	&	0.021 \\
 \hline
  \end{tabular}
  \caption{Optimal gate errors as a function of gate time.
    The optimization is performed here using the constraint
    that the population of the decaying intermediate state $| i\rangle$ be
    minimized [control functional Eq.(\ref{eq:j_stateconstraint})].
    The right column gives the gate error when the optimal pulse
    is applied and an exponential loss rate is assumed from the 
    state $| i\rangle$ (lifetime $27.7\,{\rm ns}$).
%
    }
  \label{tab:loss}
\end{table}
Table~\ref{tab:loss} reports the minimal gate error $\epsilon = 1 - F$ 
for a number of gate times $T$. Optimization was
performed for a non-local phase in the $|11\rangle$ state,
utilizing the control functional~(\ref{eq:j_stateconstraint}), i.e.,
we avoid spontaneous decay by minimizing the population of
the intermediate state, $|i\rangle=|5{\rm p}_{1/2},F=2,M_F=2\rangle$.
Other noise sources are neglected yet (see estimates in Sec.~\ref{sec:noise} 
below).
  The calculation allows for maximum Rabi frequencies
  $\Omega_{R,0} = \Omega_{B,0} = 2\pi \cdot 260\,{\rm MHz}$,
  cf. Ref.~\cite{GaetanNatPhys09},
  and assumes a detuning $\Delta = 2 \pi \cdot 600\,{\rm MHz}$
  from $| i \rangle$ (see Fig.~\ref{Fig:OneAtomHam}). A number of
  $N_r = 200$ spatial grid points is taken for the relative nuclear
  motion.
For gate durations larger than
$50\,$ns, gate errors of a few percent
are obtained. The error increases only slightly when exponential loss 
from the state $| i \rangle$ is taken into account (column ``with loss''). 
The difference between the gate error with and without
loss illustrates how well the condition of suppressing population in
$|i\rangle$ is fulfilled. 
%
For comparison, if the additional constraint
is not employed in the optimization, the gate errors increase by two
orders of magnitude to 
20\% and even more when spontaneous emission is taken into account.
The gate errors without exponential decay in $|i\rangle$
(central column of Table~\ref{tab:loss})
illustrate when excitation of the motional degree of freedom start to
play a role -- for gate durations of $70\,$ns and larger, the error
obtained without spontaneous emission is mainly due to excitation of
higher trap states. Once spontaneous emission is included, this effect
is, however, not dominant anymore. 

\begin{figure}[tbp]
  \centering
  \includegraphics*[width=0.9\linewidth]{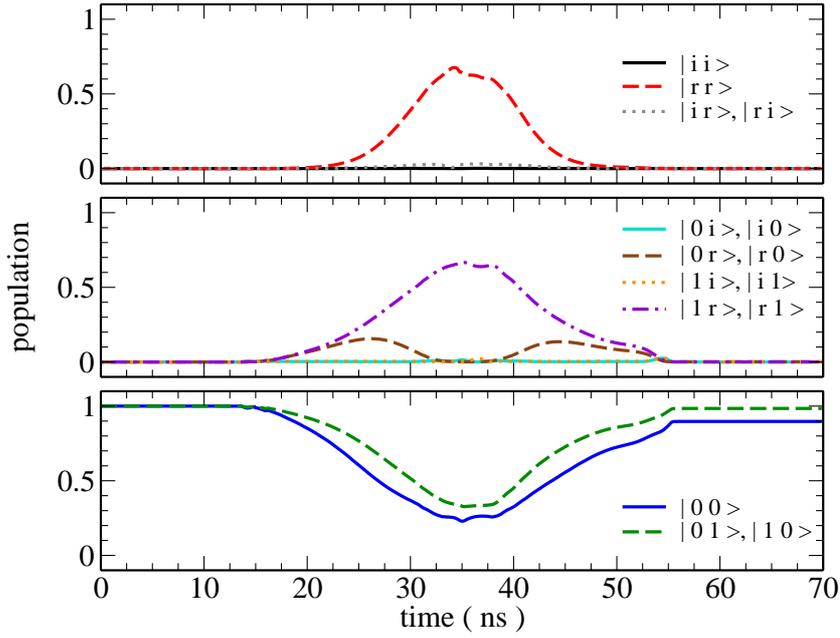}
  \caption{Minimizing the population in the unstable state $| i \rangle$:  
    population dynamics for
    the sixteen electronic states ($|11\rangle$ is not shown since the
    population is equal to one for all times). The radiative lifetime of
    the states $| i \rangle$  and $\rydbergKet$
    is described by a non-Hermitian Hamiltonian.
    }
  \label{fig:dynloss}
\end{figure}
\begin{figure}[tbp]
  \centering
  \includegraphics*[width=0.95\linewidth]{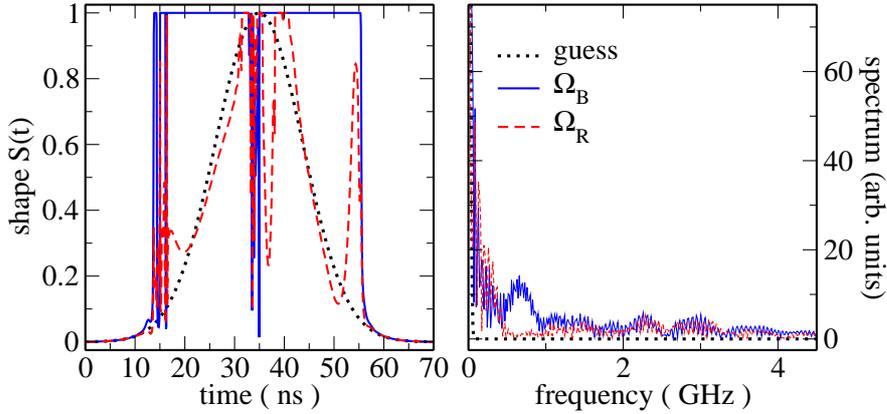}
  \caption{Optimal shapes of the laser fields and corresponding
    spectra, 
    employing the 
    constraint to minimize population of $| i \rangle$.
    }
  \label{fig:pulseloss}
\end{figure}
The population dynamics of each electronic state 
induced by the optimal fields is shown in Fig.~\ref{fig:dynloss},
demonstrating that indeed the intermediate state is
almost never populated. The overall picture suggested by the population 
dynamics is an
adiabatic transfer to and from the Rydberg state in a
double-STIRAP-like fashion. This is 
confirmed by inspecting the shapes of the optimal fields in
Fig.~\ref{fig:pulseloss}. 
The blue field connecting $|i\rangle$ and $\rydbergKet$ plays the role
of the Stokes pulse and the red field connecting $|0\rangle$ and
$|i\rangle$ that of the pump pulse in the first half of the time
interval. These roles are reversed in the second half. The middle part
of the interval, between 20$\,$ns and 50$\,$ns is crucial for
obtaining the non-local phase, as can be seen from the significant
population of the $\rrKet$ state. The dynamics of the $| 00 \rangle$
and the $| 01 \rangle$, $| 10 \rangle$ states are locked. This is due to the
fact that we optimize for a non-local phase $\chi$ in the
$|11\rangle$ state. However, the $|11\rangle$ state does not couple to the
laser fields. The non-local phase, which is given by
$\chi=\phi_{00}-\phi_{01}-\phi_{10}+\phi_{11}$~\cite{Goerz}, is thus
achieved by `coordinating' the time evolution of the remaining three
two-qubit states which can be controlled by the laser fields.

\subsection{Ignoring spontaneous emission: Rabi flop sequence}

\begin{table}[tbp]
  \centering
  \begin{tabular}{|c|l|}
    \hline
    gate time $T$ & gate error $\epsilon$ \\\hline
    20ns &	0.067 \\
    25ns &	0.051 \\
    30ns &	0.001 
    $\to$ 0.003 
    \\
    35ns &	0.001 \\
    40ns &	0.001 \\
    45ns &	0.002 \\
    50ns &	0.003 \\
    60ns &	0.003 \\
 \hline
  \end{tabular}
  \caption{
  Optimal gate errors as a function of gate time, neglecting losses
  from both the intermediate state $| i \rangle$ and the Rydberg state
  $\rydbergKet$.
  The shortest
    possible gate time to achieve a reasonable fidelity
    is then limited by the interaction in the excited state and
    the motion in the trap.
  The error increases to $\to 0.003$ (at
  $30\,{\rm ns}$) when a lifetime of $20\,\mu{\rm s}$ for the Rydberg
  level is included, consistent with the estimates of Sec.~\ref{s:cavity-QED}.
}
  \label{tab:noloss}
\end{table}
If the Rydberg state could be accessed directly, gate error $\epsilon$ and 
time $T$ would be solely determined by the interaction strength in the Rydberg
state and the trap frequency~\cite{Goerz}. We examine this scenario by
ignoring the instability of the intermediate state $| i \rangle$ and applying
the standard control functional, Eq.~(\ref{eq:j_functional}).
Table~\ref{tab:noloss} lists the minimal gate error $\epsilon$ 
neglecting the spontaneous decay from the
intermediate state for a number of gate times $T$.
Optimization was performed for a non-local phase in the $|00\rangle$
state. Inspection of Table~\ref{tab:noloss} reveals that the 
minimum gate operation time is about 30$\,$ns with a gate error
of the order of 10$^{-3}$. This limit is imposed by the interaction
strength in the Rydberg state $\rrKet$.
The fact that the gate error increases
as $T$ is enlarged, indicates that excitation of motion in the trap
starts to play a role. 
Including the trapping potential in the
calculation for the qubit
states during the gate allows motional excitation to be further 
reduced~\cite{MMMueller}.

\begin{figure}[tbp]
  \centering
  \includegraphics*[width=0.9\linewidth]{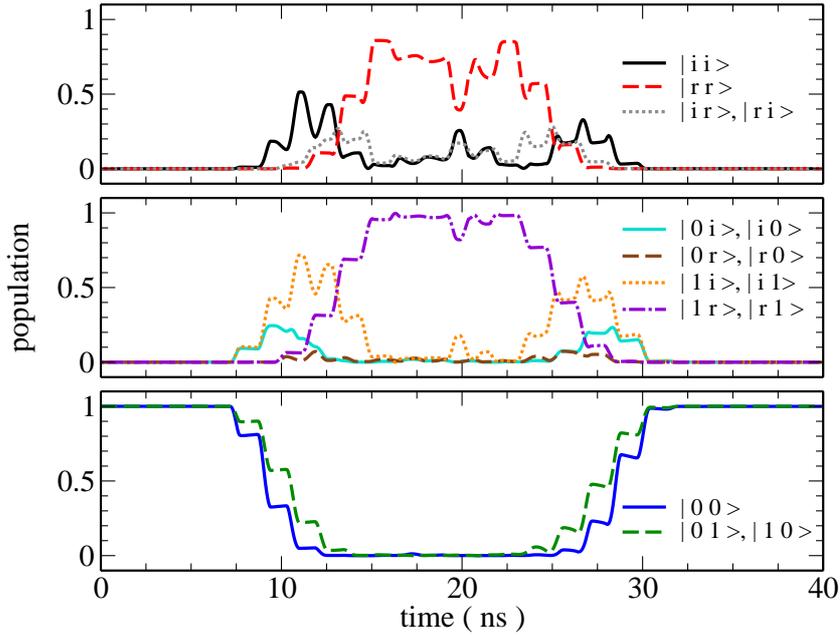}
  \caption{Two-photon excitation scheme treating the intermediate state 
  $| i \rangle$ as stable: population dynamics for
    the sixteen electronic states. ($|11\rangle$ is not shown since the
    population is equal to one for all times.)}
  \label{fig:dynnoloss}
\end{figure}
\begin{figure}[tbp]
  \centering
  \includegraphics*[width=0.95\linewidth]{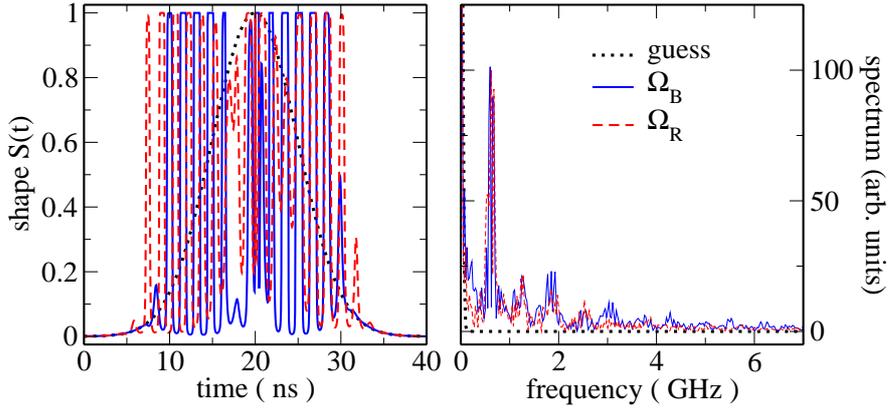}
  \caption{Optimal shapes of the laser fields and corresponding
    spectra, assuming the intermediate state $| i \rangle$ to be stable.
}
  \label{fig:pulsenoloss}
\end{figure}
The population dynamics of each electronic state 
induced by the optimal fields shown in Fig.~\ref{fig:dynnoloss} are
clearly non-adiabatic. The intermediate state $|i\rangle$ is
significantly populated at intermediate times 
since the optimization algorithm does not `know' that  $|i\rangle$ 
 corresponds
to a loss channel. The population of the Rydberg state in the middle
of the time interval is larger in Fig.~\ref{fig:dynnoloss} than in
Fig.~\ref{fig:dynloss}. This explains why a smaller gate duration can
be achieved. The optimal pulse shapes and spectra are shown in
Fig.~\ref{fig:pulsenoloss}. The observed sequence of fast switches
of both red and blue laser fields pumps population in a ladder-like
fashion with a time constant equal to the duration of a $\pi$-pulse
at the maximally allowed Rabi frequency. Correspondingly, we observe
stronger sidebands in the spectra of the optimal pulses (compare
Figs.~\ref{fig:pulseloss} and~\ref{fig:pulsenoloss}).

In summary, we find minimum gate durations of about 30$\,$ns to
50$\,$ns for a high-fidelity implementation of the Rydberg gate. 
Shorter gate operation times might be possible by employing higher
lying Rydberg states which exhibit a stronger interaction. However,
higher states will also be more sensitive to stray fields. 
While our model includes all relevant degrees of freedom and
spontaneous emission from the excited states, noise sources that are
inevitably present in any experimental setup need to be analysed
in order to gain a full understanding of what fidelities can be achieved in
an experiment based on optimized pulse shapes. The influence of noise
sources is expected to be particularly detrimental for miniaturized
setups such as an atom chip.

\section{Influence of noise}
\label{sec:noise}

The hyperfine ground-state levels $| 0 \rangle$, $| 1 \rangle$,
as defined in Sec.~\ref{subsec:rydberg},
provide a qubit robust with respect to noise, as has been discussed 
elsewhere \cite{FolmanAMOP02,FortaghRMP07,ReichelVuleticBook}
and demonstrated experimentally \cite{TreutleinPRL04}. We therefore focus here 
on the specific sensitivity of the Rydberg state $\rydbergKet$ to static
and fluctuating fields typical for an atom chip environment. Since the
Rydberg levels are populated only over the duration of the laser pulse
(some $10\,{\rm ns}$) and their angular momenta are relatively small
($j \le 5/2$),
we do not have to consider magnetic field noise
that takes effect only on a time scale of $100\,{\rm ms}$ or 
longer \cite{ReichelVuleticBook}.

\subsection{Surface impurities: DC Stark shifts}
\label{s:static-fields}


Previous work has shown that the surfaces of atom chips get contaminated
with adsorbed atoms that lead to randomly placed charges or 
dipoles~\cite{McGuirkPRA04}. As a typical scenario, consider an alkali atom
on a metal that gives off its electron into the surface so that an upright
dipole moment remains. At a distance $z \approx 10\,\mu{\rm m}$ above
it, this single impurity creates a tiny electric field of the order of
\begin{equation}
	E_{\rm imp}( z ) \sim \frac{ 2 e a_B }{ 4\pi\varepsilon_0 z^3 } =
	0.15\,\frac{ \mathrm{mV} }{ \mathrm{m} }\,,
	\label{eq:estimate-dipole-field}
\end{equation}
where $a_B$ is the Bohr radius. But if $10^4$ adatoms are distributed
over an area of $(10\,\mu{\rm m})^2$, (corresponding to an average distance
$\sim 100 \, {\rm nm}$, i.e., much less than a monolayer), their electric fields
roughly add up to ${\cal O}( 1\, {\rm V}/{\rm m} )$.  The
corresponding Stark-Hamiltonian,
\begin{equation}
	H_{\rm S} = - \Op{\mu} \cdot {\bf E}_{\rm imp}( z )\,,
	\label{eq:DC-Stark-coupling-to-impurity-field}
\end{equation}
has matrix elements of the order $e a_B n^2 E_{\rm imp}(z) 
\approx 6.6\,{\rm MHz}$ for
$n = 58$, just one order of magnitude below the Rydberg interaction
Hamiltonian~(\ref{eq:Hint}). 
The experiments of Ref.~\cite{McGuirkPRA04} have actually detected
electric fields up to $1\,{\rm kV}/{\rm m}$ at a distance of about 
$10\,\mu{\rm m}$. On an insulating surface, charges can be trapped
and the fields would even be stronger. Note that the threshold for
field ionization  is of the order
of $5\,{\rm kV}/{\rm m}$ for the $n = 58$ level \cite{GallagherRPP88}.

\begin{figure}[tb]
\includegraphics*[width=0.5\linewidth]{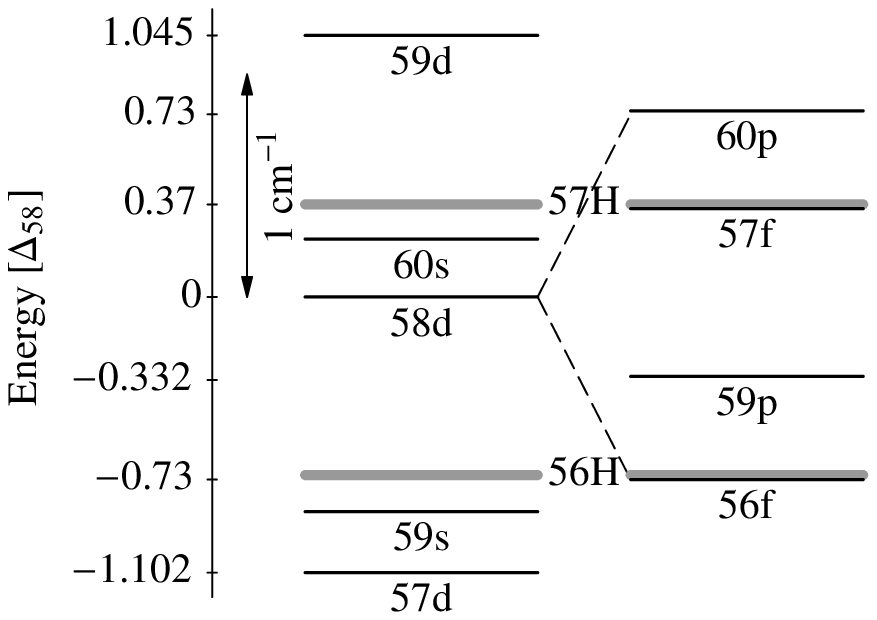}
\includegraphics*[width=0.5\linewidth]{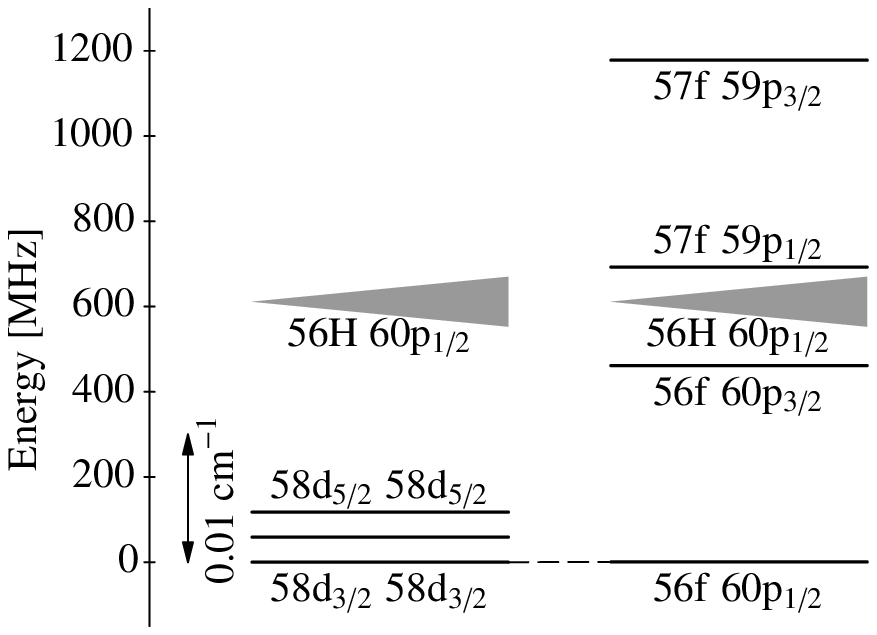}
\caption[]{(left)
Energy levels near the Rydberg state 
$\rydbergKet = | 58{\rm d}_{3/2}\rangle$. The strong Rydberg interaction
arises from the near degeneracy of $|r r\rangle$ with the two-atom 
level $56 {\rm f}_{5/2}\,60 {\rm p}_{1/2}$ (``F\"orster resonance''
\cite{WalkerJPB05}, dashed lines). 
The fine structure is
not resolved on the scale $\Delta_{58} = 1 / 58^3 \, {\rm a.u.} =
33.7\,{\rm GHz}$.
 The thick gray lines (label $n{\rm H}$) 
give the hydrogen-like
state manifold beyond the $n {\rm f}$ states (angular momenta $l > 3$).
The left and right columns show even and odd orbital angular momentum
states.\\
(right)
Energy levels of the two-atom system. The levels in the F\"orster resonance 
are connected by the dashed line (detuning $\approx 7\,{\rm MHz}$ 
\cite{GaetanNatPhys09}). At a distance $r_0 \approx 4\,\mu{\rm m}$,
the two levels hybridize and split by $\approx 50\,{\rm MHz}$, as described
by the two-atom Hamiltonian~(\ref{eq:Hint}).
Note the difference in scale: the fine structure
of the ${\rm p}$ and ${\rm d}$ levels is resolved (but not for the ${\rm f}$
levels). A quantum defect of order $10^{-2}$ splits off
the level $56{\rm f}$ from the hydrogen-like manifold $56{\rm H} = 
56{\rm g},{\rm h}, \ldots$.
The gray triangles illustrate the linear Stark splitting
of $56{\rm H}$ in a weak static electric field between $0$ and 
$1{\rm V}/{\rm m}$ (shifts up to $\pm 60\,{\rm MHz}$).
At slightly higher fields, also the levels 
$57{\rm f}\,59{\rm p}_{1/2}$ and 
$56{\rm f}\,60{\rm p}_{3/2}$ would be pushed away (avoided crossings).
The quadratic Stark shift of the other levels is not visible on this scale.
\\
The energy levels are calculated from quantum defect data collected
in Refs.~\cite{LorenzenPS83,LiPRA03}. A precise localization of the F\"orster resonance
(not attempted here) would require knowledge of the quantum defects 
at the $10^{-4}$ level.
}
\label{fig:level-scheme}
\end{figure}
The Rydberg level considered here, $\rydbergKet =
| n {\rm d}_{3/2} \rangle$ ($n = 58$), is actually ``protected'' from a
linear Stark shift because its quantum defect 
($\delta_{\rm d} \approx 1.34$) lifts the degeneracy with
the opposite parity states $| n {\rm p}, {\rm f}, \ldots \rangle$,
by an energy splitting of the order of $\Delta_n = 1/n^3\,{\rm a.u.} \approx 
1.125\,{\rm cm}^{-1} = 33.7\, {\rm GHz}$ (see 
Fig.~\ref{fig:level-scheme}~(left)). For this reason, the Stark shift is
quadratic in the field and inversely proportional to the detuning 
($< \Delta_n$) from the nearest level with opposite parity. 
The nearest levels $57{\rm f}$ and $59{\rm p}$ give quadratic Stark shifts
that partially cancel each other, 
leaving a polarizability for $58 {\rm d}$
of the order of $50\,{\rm kHz} ({\rm V}/{\rm m})^{-2}$. 
The $56{\rm f}$ state shows a much
larger polarizability because the hydrogen manifold $56{\rm H} = 
56{\rm g}, {\rm h}, \ldots$ is only about 
$0.02\,{\rm cm}^{-1}$ ($\approx 600\,{\rm MHz}$) away 
(Fig.~\ref{fig:level-scheme}~(right)) and its influence is not cancelled by
another level. We estimate a quadratic
Stark effect of the order of $2\ldots3\,{\rm MHz} ({\rm V}/{\rm m})^{-2}$.

This means that
fields above $1\,{\rm V}/{\rm m}$ start to detune the 
so-called F\"orster resonance between the two-atom 
states
$| 58 {\rm d}_{3/2}\,58 {\rm d}_{3/2} \rangle$ and
$| 56 {\rm f}_{5/2}\,60 {\rm p}_{1/2} \rangle$ whose
energy mismatch is only
$7\,{\rm MHz}\approx 2.3 \cdot 10^{-4} \,{\rm cm}^{-1}$ 
(see Fig.~\ref{fig:level-scheme}~(right) and
Refs.~\cite{WalkerJPB05,GaetanNatPhys09}). One then loses the
strong $1/r^3$ scaling of the Rydberg interaction that turns into the
weaker $1/r^6$ van der Waals scaling. In addition, at the level of 
$10\,{\rm V}/{\rm m}$, 
the linear DC Stark shift of the hydrogen-like state manifold 
$| 56 {\rm H}\rangle$ exceeds $\approx 600\,{\rm MHz}$.
By an avoided crossing, the state $| 56 {\rm f}_{5/2}\rangle$
is then pushed down, and the F\"orster resonance is detuned.



\subsection{Fluctuating fields: dephasing}

\begin{figure}[bt]
\centerline{
\includegraphics*[width=0.8\textwidth]{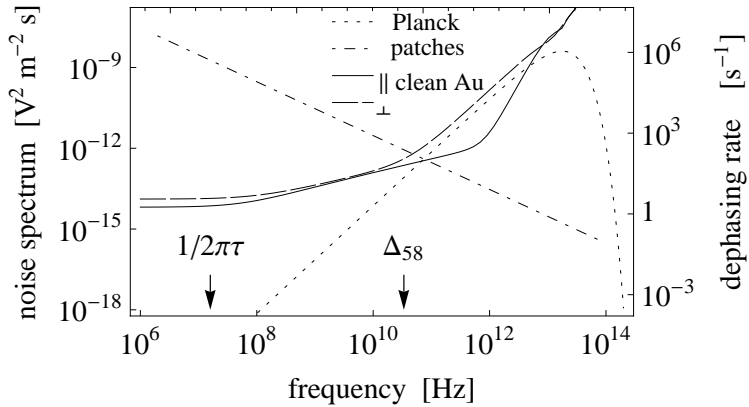}
}
\caption[]{Electric field noise spectrum (in $({\rm V}/{\rm m})^2/{\rm Hz}$)
due to patch charge fluctuations
(dot-dashed) and due to Johnson-Nyquist noise from a gold half-space
(solid and dashed lines).
Distance fixed to $z = 10\,\mu{\rm m}$, temperature 300\,K.
Patch fluctuations are calculated from Ref.~\cite{DubessyPRA09} and
extrapolate ion trap data (in the MHz range), assuming a $1/\omega$ 
scaling. 
The Johnson-Nyquist noise is calculated along the lines of 
Ref.~\cite{HenkelEPL99}. It includes blackbody radiation and free-space
vacuum fluctuations and changes from a $\omega^{1/2}$ into a
$\omega^2$ power law near the typical Rydberg transition energy
$\Delta_{58}$ (arrow). Solid (dashed) lines are for fields parallel (perpendicular)
to the surface, respectively. The right scale gives the expected dephasing
rate for a static dipole moment ${\cal O}( 200 \, e a_B )$ and a white
spectrum. The arrow at
$1/2\pi\tau$ marks the Fourier-limited band width for quasi-static noise over
an effective interaction time $\tau = 10\,{\rm ns}$. The contribution of
quantum fluctuations was subtracted in the Planck spectrum (dotted curve).
}
\label{fig:dephasing-spectrum}
\end{figure}
The one- and two-atom Rydberg states that we are considering do
not feature a permanent electric dipole moment. This is due to the quantum
defects that split them off the hydrogen-like manifolds of higher
angular momentum states ($l > 3$). In weak fields, their Stark effect is
therefore quadratic. We consider here the beating between a fluctuating 
field and a static impurity field $E_{\rm imp}( z )$
at the level of $1\,{\rm V}/{\rm m}$, small enough not to perturb 
the dipole-dipole interactions, see previous section.
The mixing of states with opposite parity is linear in the impurity field 
and translates into a dipole moment of the order of
$d_R = \alpha_R E_{\rm imp}( z )$ where $\alpha_R$ is the quasi-static
polarizability. For the level $58{\rm d}_{3/2}$, the contributions to
nearby levels above and below partially cancel, leading to the relatively 
small value $|d_R| \sim 4\,e a_B$. The level $56{\rm f}_{5/2}$ is
much more polarizable, as found above, and $d_R \sim 200\, e a_B$.
Note that this is 
still much smaller than expected from the average size of a Rydberg 
atom ($\approx a_B n^{2}$).
%
The fluctuating phase shift due to the Rydberg polarizability
$\alpha_R$ is then given by 
\begin{equation}
	\Delta \phi(\tau) = - \int\limits_{t_1}^{t_2}\!{\rm d}t\,
	\frac{ \alpha_R }{ \hbar }
	{\bf E}_{\rm imp}( z ) \cdot
	{\bf E}_{\rm fluct}( z, t )
	\,,
	\label{eq:def-phase-shift}
\end{equation}
where the time integral is evaluated over that part of the pulse
that the atom effectively spends in the Rydberg level (see 
Figs.~\ref{fig:dynloss}, \ref{fig:dynnoloss}).

Let us assume an electric field noise spectrum with a scaling
\begin{equation}
	S_E( z, \omega ) = \frac{ S_E( z_0, \omega_0 ) }{ 
	( z / z_0 )^4 }
	\left( \frac{ \omega_0 }{ \omega } \right)^\beta
	\label{eq:power-law-scaling-for electric-noise}
\end{equation}
that arises from ``patch charge'' fluctuations, as observed
in experiments with miniaturized ion
traps~\cite{DubessyPRA09,TurchettePRA00,LabaziewiczPRL08}  
(exponent $\beta \approx 0.7 \ldots 1$). The $1/z^4$ scaling 
actually only holds at heights $z$ beyond a characteristic length scale
for the patch size. This effect is taken into account in the 
plots shown, using the model of Ref.~\cite{DubessyPRA09}.
%
The thermal electric fields that originate from the motion
of charges in the chip material (Johnson-Nyquist noise) have a lower
noise spectrum compared to the patch charge model of 
Eq.~(\ref{eq:power-law-scaling-for electric-noise})~\cite{HenkelEPL99,TurchettePRA00,LeibrandtQIC07}, at least in the
low-frequency range (up to a few MHz) relevant for dephasing.
This is illustrated
in Fig.~\ref{fig:dephasing-spectrum} where the two noise spectra (at 
fixed distance $10\,\mu{\rm m}$) are plotted vs frequency. 

The dephasing of the Rydberg state is now estimated by calculating
the variance of the phase shift~(\ref{eq:def-phase-shift}). 
Provided the interaction time $\tau=t_2-t_1$ is large compared
to the noise correlation time, the variance increases like
\begin{equation}
	\overline{ \Delta \phi^2( \tau ) } \approx
	\tau 
	 \frac{ d_R^2 S_E( z_0, 1/\tau ) }{ 2 \hbar^2 
	 (z / z_0)^4 }
	 \frac{ 1 }{ \cos(\pi \beta / 2) 
		 \Gamma(2 + \beta) }\,,
	\label{eq:estimate-dephasing}
\end{equation}
where the noise spectrum is evaluated at $\omega = 1/\tau$,
roughly the Fourier-limited band width of the pulse, and 
$\Gamma( \cdot )$ is Euler's gamma function.
For room-temperature microscopic ion 
traps~\cite{LabaziewiczPRL08,DubessyPRA09,DeslauriersPRL06,EpsteinPRA07}, 
noise levels of $S_E( 75\,\mu{\rm m}, 
1\,{\rm MHz} ) \sim 10^{-11}\,({\rm V}/{\rm m})^2 / {\rm Hz}$ 
and $\beta \approx 0.7 \ldots 0.8$ are typical. 
Note that a low-frequency cutoff must be applied
for pure $1/f$ noise. Taking an effective time of
$\tau \approx 20\,{\rm ns}$ spent in the Rydberg level, 
one gets at $z = 10\,\mu{\rm m}$ a dephasing rate of 
\begin{equation}
	D_\phi(56{\rm f}) = \overline{ \Delta \phi^2( t ) } / \tau
	= {\cal O}( 10^6\,{\rm s}^{ - 1 } )
	\label{eq:estimate-dephasing-rate-worst-case}
\end{equation}
for the highly polarizable Rydberg state. This affects in particular the
two-Rydberg state via its admixture of $56{\rm f} \, 60{\rm p}$
(Fig.~\ref{fig:level-scheme}).
For the phase gate, we get a sizable
decoherence error $\epsilon \approx 1 - \langle {\rm e}^{ {\rm i}
\Delta \phi( \tau ) } \rangle \approx 
\frac12 D_\phi \tau \sim 1\%$
from the uncertainty in the two-atom phase $\phi_{11}$. We recall
that this number scales quadratically with the impurity field, assumed  
here to be $1\,{\rm V}/{\rm m}$.
The other 
Rydberg levels involved show a much smaller dephasing, in particular
$D_\phi(58{\rm d}) = {\cal O}( 400\,{\rm s}^{ - 1 } )$.
This is insignificant over the effective pulse duration $\tau$, and therefore
the one-atom phases $\phi_{01}$, $\phi_{10}$ are not compromised.

\begin{figure}[tb]
\includegraphics*[width=0.7\textwidth]{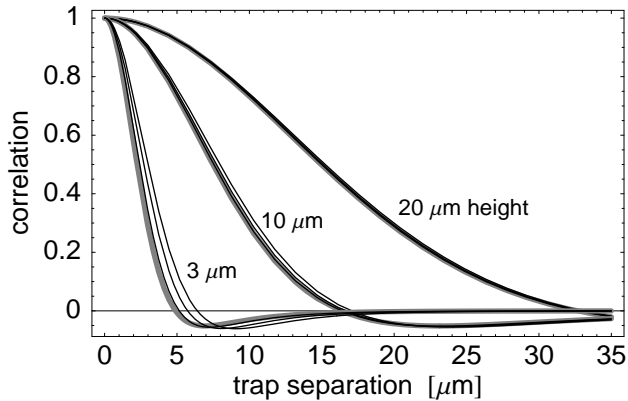}
\caption[]{Normalized cross-correlation of electric field noise due to
patch charge fluctuations, evaluated at three
different positions and normalized to the noise spectrum in one trap. If the
correlation is weak, the relative phase of the Rydberg state in the two traps
is randomized on the same time scale as the common phase.
The correlations are calculated by generalizing the model of 
Ref.~\cite{DubessyPRA09}. Any group of three curves corresponds to
patch correlation lengths $0.5$, $1$, and $1.5\,\mu{\rm m}$.}
\label{fig:lateral-correlation}
\end{figure}

One may ask the question to what extent the fields seen by the two atoms are
correlated. The phase gate misses its target already, of course, if there are 
fluctuations 
that are common to both atoms. A differential phase would mix the even and
odd states 
$| 56 {\rm f}_{5/2} \, 60{\rm p}_{1/2} \rangle \,\pm \,
| 60{\rm p}_{1/2} \, 56 {\rm f}_{5/2} \rangle$ that are involved in the
F\"orster resonance. This can be quantified from the cross-correlation spectrum
of the patch fields, $S_E( A, B; \omega )$, where $A$ and $B$ are the 
positions of the two traps. 
In Fig.~\ref{fig:lateral-correlation}, we plot the normalized cross-correlation
for two atoms at the same distance $z$ from the chip, but laterally separated 
by a length $r_0$. The two atoms are subject to the same noise (no differential 
dephasing) if the normalized correlation is unity. One notes that the fields
decorrelate on a scale given by the height above the surface. The correlation
is still quite strong for the parameters we considered here, i.e. a separation
$r_0 = 4\,\mu{\rm m}$ and height $10 \,\mu{\rm m}$.

\subsection{Thermal radiation: lifetime and AC Stark shift}
\label{s:cavity-QED}



\begin{figure}[bt]
\centerline{
\includegraphics*[width=0.8\textwidth]{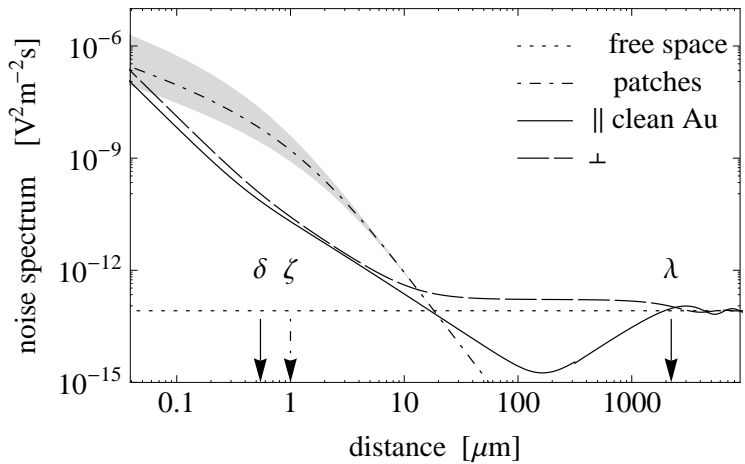}
}
\caption[]{
Electric field noise spectra (in $({\rm V}/{\rm m})^2/{\rm Hz}$)
vs distance from the chip surface, evaluated at the level splitting
$\Delta_{58}$ typical for Rydberg transitions, and 300 K. 
The patch charge spectrum is extrapolated to this frequency assuming
a $1/\omega$ scaling; the patch correlation length is in the range
$\zeta = 1 \pm 0.5\, \mu{\rm m}$ (shaded area). The solid and dashed
curves ``clean Au'' give the noise spectrum of thermal near field
radiation, the far field limit is shown by the dotted line. The skin depth 
$\delta$ separates two power laws~\cite{HenkelEPL99}, and for
$z \gtrsim \lambda/2$, the spectrum oscillates due to interference
by reflection from the surface. In this limit, a perfect conductor approach
yields good agreement. Within the wide plateau betwen $10\,\mu{\rm m}$ 
and  $1\,{\rm mm}$, a dipole perpendicular to the surface
($\perp$) is subject to near field noise about five times as strong as 
the free space (Planck) level.}
\label{fig:spectrum-distance}
\end{figure}
The radiative lifetime of an emitter is strongly modified in the vicinity
of a macroscopic body, as illustrated by the Purcell effect~\cite{PurcellPR46}.
In addition, thermal radiation plays a significant role because the Bohr 
frequencies of Rydberg atoms are low, typically $\Delta_n \ll
k_B T$. It is well known 
that this reduces the lifetime and coherence time of trapped particles, even to the
level that magnetic dipole transitions become relevant~\cite{ReichelVuleticBook}.
We estimate here surface-enhanced radiative loss 
and discuss that 
despite the large polarizabilities, thermal radiation gives rise to
AC Stark shifts that are overall small, leaving only the zero-temperature
van der Waals shift. A similar question has been addressed in
Refs.~\cite{CrossePRA10,EllingsenPRL10}. 

Spontaneous emission in free space is 
dominated by the decay into the lowest states because of the cubic
scaling of the vacuum field spectrum.
This leads to a lifetime of the Rydberg state of $210\,\mu$s at $T = 0$ 
\cite{TheodosiouPRA84} and a decay probability, i.e., gate error, of
the order of $10^{-4}$ given the Rydberg excitation time 
$\tau \approx 20\,{\rm ns}$. We therefore
need to check that this error does not increase significantly
in an atom chip environment.
The contribution of blackbody radiation reduces the lifetime significantly 
(absorption and emission), to a total figure of roughly $90\,\mu{\rm s}$ 
for the Rydberg level $58d$ where the effective quantum number is 
$n^* \approx 58 - 1.34$~\cite{TheodosiouPRA84}.

The chip surface enhances the mode density of the electromagnetic 
field~\cite{CrossePRA10,WyliePRA84,FailachePRL01}. This leads to a 
different scenario, however, for transitions with large or small Bohr
frequencies. 
For the decay into low states, the resonant wavelengths are typically 
in the visible and near UV, small compared to the atom-surface distance. 
These field modes form a 
shallow interference pattern due to reflection at the surface, enhancing
or suppressing the decay by roughly a factor of $2$. 
Destructive interference can be used to suppress certain decay channels, 
as suggested in Ref.~\cite{HyafilPRL04} in an application of the Purcell
effect \cite{PurcellPR46}. 
%

At smaller Bohr frequencies, thermally stimulated emission
and absorption are enhanced much more strongly in
the near field. This can be calculated from the
approach of Wylie and Sipe \cite{WyliePRA84,CrossePRA10}. 
The rate for a transition $r \to s$ is given by
\begin{equation}
	\gamma_{r \to s} = \frac{ 2 [ 1 + \bar n( \omega_{rs} ) ]}{ \hbar }
	\sum_{kl}
	\langle s | \mu_k \rydbergKet \langle s | \mu_l \rydbergKet^*
	\,\mathfrak{Im}\left[ G_{kl}( z; \omega_{rs} )\right]\,,
	\label{eq:decay-rate}
\end{equation}
where the Bohr frequency is $\hbar \omega_{rs} = E_r - E_s$,
and the thermal occupation number 
$\bar n( \omega_{rs} )$ is evaluated at temperature $T$. 
The matrix elements of the dipole operator $\Op{\mu}$ are written
in Cartesian components, and $G_{kl}$ is the
electromagnetic Green tensor at the position $z$ of the atom. Note
that we normalize
it such that $\varepsilon_0 G_{kl}$ has units of inverse volume.
This formula also
describes the absorption rate of thermal photons 
($\omega_{rs} = - \omega_{sr} < 0$) because
\begin{equation}
	[ 1 + \bar n( -\omega_{sr} ) ]
	\, \mathfrak{Im}\left[ G_{kl}( z ; -\omega_{sr} )\right]
	=
	\bar n( \omega_{sr} )
	\, \mathfrak{Im} \left[G_{kl}( z ; \omega_{sr} )\right]
	\,.
	\label{eq:excitation-rate}
\end{equation}
We get the total decay rate $\gamma_r$ 
by summing Eq.~(\ref{eq:decay-rate}) over the final states $|s\rangle$.
Fig.~\ref{fig:dephasing-spectrum} can be taken as an illustration of the
terms in this sum because the electric field noise spectrum is proportional
to $\mathfrak{Im} \left[G_{kl}( z; \omega_{rs} )\right] [1 + \bar n( \omega_{rs})]$.
The Green tensor is calculated for a gold surface using the formulas
of Ref.~\cite{WyliePRA84}. The dependence on distance is shown in
Fig.~\ref{fig:spectrum-distance}.
Compared to
free space (Planck spectrum, dashed line), transitions among Rydberg
levels are significantly enhanced at short distances, while the rates
oscillate in the opposite limit (distance comparable to the
transition wavelength)
due to the interference pattern mentioned above.
Note the quite strong destructive interference for a transition 
dipole parallel to the surface (solid curve) which can be understood from
the image dipole at a perfectly conducting surface.

The transition rates scale with the electric dipole matrix elements of the
Rydberg levels. We note that for a Bohr frequency 
$\hbar |\omega_{rs}| \gg \Delta_n$, the matrix elements are
much smaller because a kind of radial selection rule suppresses
changes in the principal quantum 
number by more than a few units (see also 
Refs.~\cite{GallagherRPP88,CrossePRA10}). 
For a typical final state among adjacent Rydberg levels, for example
$\Delta n^* \lesssim 2$, 
we find an enhancement of the  transition rate by a factor
${\cal O}( 5 )$ at $z = 10\,\mu{\rm m}$ compared to free space. 
The corresponding lifetime 
is reduced from $\sim 150\,\mu{\rm s}$ (free space) to 
$\sim 30\,\mu{\rm s}$,
estimating the matrix element by $e a_B n^{*2}$.
We note that the rate for this generic pair
of levels essentially
exhausts the thermal decay rate summed over all final states, including
photoionization \cite{TheodosiouPRA84}. We therefore expect a total 
lifetime somewhat
below the $\sim 30\,\mu{\rm s}$ figure estimated above from a single
bound-bound transition. While this is still three orders of magnitude
longer than the Rydberg excitation pulse, it may pose a serious challenge
to gate errors below the fault tolerance threshold of $10^{-4}$. 
As a
preliminary check, we have performed (see Table~\ref{tab:noloss})
a calculation of the phase gate error when a finite lifetime 
$20\,\mu{\rm s}$ of the Rydberg level is included: an increase of the 
order of $10^{-3}$ is indeed found.



The radiation field also induces a van der Waals--Casimir--Polder shift
on the Rydberg levels. We can estimate this in the London limit (transition
wavelength large compared to the atom-surface distance) since the
Rydberg spacing $\Delta_{58}$ corresponds to wavelengths in the centimeter
range,
\begin{equation}
	V_{\rm vdW}( z ) = - \frac{ \rydbergBra \Op{\mu}_x^2 
		+ \Op{\mu}_y^2 + 2 \Op{\mu}_z^2 \rydbergKet 
		}{ 8 \pi \varepsilon_0 (2 z)^3 }\,.
	\label{eq:van-der-Waals-shift}
\end{equation}
The expectation value of the squared dipole is of the order of 
$\frac52 (e a_B n^{*2})^2$ (see, e.g., Ref.~\cite{CourtoisPRA96}) 
and gives a shift $\approx 1.6\,{\rm MHz}$ at $10\,\mu{\rm m}$,
consistent with the findings of Ref.~\cite{CrossePRA10} where 
the electric quadrupole contribution is analyzed as well.
This is not far from the dipole-dipole interaction (as it must from the
scaling), but still small enough not to perturb it. The level shift changes
only weakly across the levels shown in Fig.~\ref{fig:level-scheme} and
does not induce significant detunings. Its
main impact is therefore to pull the Rydberg atoms towards the chip 
during the pulse. This effect which excites motional states in the trap, 
could be compensated for by the exciting laser pulse in a 
similar way as the momentum exchange between the two Rydberg atoms 
(see Sec.~\ref{subsec:gatetime}).

We finally turn to the question how thermal radiation is shifting the 
Rydberg levels. This could be significant since even room-temperature
blackbody radiation produces a sizable electric field above 
$100\,{\rm V}/{\rm m}$, albeit over a wide frequency range.
It has been pointed out that 
the Casimir--Polder potential is essentially
temperature-independent,
due to cancellations
between different transitions, on the one hand, and virtual and real photon
exchange, on the other
\cite{EllingsenPRL10}. This holds provided the 
typical transition wavelengths are large compared to the atom-surface 
distance, which is indeed the case for a Rydberg atom.
%
We have checked that this result can be understood in a simple way 
starting from the dynamical polarizability of a free electron
and integrating over the thermal radiation spectrum.
For an analysis at zero temperature, see Ref.~\cite{EberleinPRL04} and
references therein.
The Planck spectrum gives a free-space level shift,
common to all weakly bound Rydberg levels, of the order of
$\alpha_{\rm fs} (k_B T)^2 / (m_e c^2) = {\cal O}( 2\,{\rm kHz} )$,
where $\alpha_{\rm fs}$
is the fine structure constant and $m_e c^2$ the electron's rest energy.
This is 
consistent with the value quoted in Ref.~\cite{GallagherRPP88}.
Near the chip surface, the shift is modified by a factor $\lambda_T / z$,
where $\lambda_T = \hbar c / k_B T
\approx 7.6\,\mu{\rm m}$ is Wien's thermal wavelength. The result
is still negligible on the energy scale set by the Rydberg dipole-dipole
interaction, and we provide a detailed discussion elsewhere.


\section{Conclusions}
\label{sec:concl}

We have used optimal control theory to determine the shortest gate
duration for a controlled phase gate based on resonant excitation of
neutral rubidium atoms to Rydberg levels that show a long-range 
dipole-dipole interaction. The parameters were chosen to be similar to
those of the experiment reported in Ref.~\cite{GaetanNatPhys09}, in
particular a distance 
between the atoms of $4\,\mu$m, a trapping frequency of
$\omega/2\pi \approx 276\,\mathrm{kHz}$, 
and near-resonant two-photon interaction to the Rydberg level via the
$5p_{1/2}$ level. It turns out that in this setting the fidelity of
the controlled phase gate is limited to a few percent by the strong
decay from the intermediate level, despite suppressing the population
of the intermediate level as much as possible. 
The optimal pulses correspond to STIRAP-like transitions to and
from the Rydberg state, i.e. an adiabatic solution to the control
problem is found. Neglecting spontaneous decay from the
intermediate level 
(which would correspond to a different excitation
scheme to the Rydberg level), 
gate errors of the order of $10^{-3}$ are obtained
for $T=30\,$ns and larger. 
This lower limit on duration is essentially due to the 
interaction strength of the two atoms in the Rydberg state. 
The control pulses induce a sequence of Rabi flops, and the ensuing
nuclear dynamics is strongly non-adiabatic. 
Further reduction of the gate error to values below the fault tolerance
threshold of $10^{-4}$ requires the calculation to include the 
trapping potential for the qubit states 
during the gate. Otherwise, the motional state of the atoms at the end
of the gate is not well-defined to this precision, and higher lying
trap states get excited. 

These conclusions hold in a general setting where our parameters for
trap separation and frequency are applicable. In order to estimate
whether a Rydberg phase gate can be implemented on an atom chip to yield a
universal quantum computer in a scalable setting, all relevant noise
sources specific to the chip environment need to be considered.
To summarize the estimates on sensitivity to noise due to the chip
environment, we have found  
serious issues for Rydberg atoms held at a distance of the order of
$10\,\mu{\rm m}$, due to linear and quadratic Stark shifts. 
The reason is the contamination by impurity atoms of the chip 
surface at densities
higher than $100\,\mu{\rm m}^{-2}$, still much less than a 
monolayer. This would create, above a metallic surface, electric fields
exceeding ${\cal O}( 1\,{\rm V}/{\rm m} )$, and reduce
significantly the strong dipole-dipole interaction between the
Rydberg atoms. This slows down the quantum gate, and a sizable
dephasing rate arises from 
the beating between stray fields and fluctuating patch potentials,
in particular for the highly polarizable 
$56{\rm f}$ state. Operating the chip at lower temperatures
would reduce the patch charge noise \cite{LabaziewiczPRL08,DeslauriersPRL06}.
The radiative decay of the Rydberg state
is also enhanced by nearly an order of magnitude
compared to free space at zero temperature: we estimate a lifetime
in the $20\ldots 30\,\mu{\rm s}$ range. As it stands, this could provide 
a fundamental lower limit around $10^{-3}$ to the gate error. Possible
improvements may exploit a Purcell effect to suppress, by destructive
interference, radiative transitions for certain orientations of the transition 
dipole, 
similar to the suggestion 
in Ref.~\cite{HyafilPRL04}. This may be achieved with suitably polarized laser 
pulses. 
We therefore estimate that reaching errors below the fault
tolerance threshold for a Rydberg phase gate on an atom chip 
is challenging but possible.

\begin{acknowledgements}
We would like to thank Ron Folman for stimulating this work. 
Financial support from the Deutsche Forschungsgemeinschaft (Ko 2301/2),
the German-Israeli Foundation for Scientific Cooperation (982-192.14/2007),
and the European Union (Integrated Project AQUTE)
is gratefully acknowledged.
\end{acknowledgements}


\end{document}